\documentclass[graybox]{svmult}

\usepackage{mathptmx}       % selects Times Roman as basic font
\usepackage{helvet}         % selects Helvetica as sans-serif font
\usepackage{courier}        % selects Courier as typewriter font
\usepackage{type1cm}        % activate if the above 3 fonts are
\usepackage{makeidx}         % allows index generation
\usepackage{graphicx}        % standard LaTeX graphics tool
\usepackage{multicol}        % used for the two-column index
\usepackage[bottom]{footmisc}% places footnotes at page bottom

\usepackage{axodraw}

\makeindex             % used for the subject index
\begin{document}

\newcommand{\Aslash}{A \! \! \! /}
\newcommand{\Dslash}{D \! \! \! \! \! /}
\newcommand{\kslash}{k \! \! \! /}
\newcommand{\pslash}{p \! \! \! \! /}
\newcommand{\xslash}{x \! \! \! /}
\newcommand{\partialslash}{\partial \! \! \! /}
\newcommand{\half}{\mbox{\small{$\frac{1}{2}$}}}
\newcommand{\Nc}{N_{\!c}}
\newcommand{\Nf}{N_{\!f}}
\newcommand{\MSbar}{\overline{\mbox{MS}}}

\title*{Conformal methods for massless Feynman integrals and large $\Nf$ 
methods}
\titlerunning{Conformal Methods}
\author{J.A. Gracey}
\authorrunning{J.A. Gracey} 
%\institute{J.A. Gracey, \at Theoretical Physics Division, Department of 
\institute{J.A. Gracey, Theoretical Physics Division, Department of 
Mathematical Sciences, University of Liverpool, P.O. Box 147, Liverpool, 
L69 3BX, United Kingdom \email{gracey@liv.ac.uk} \\
Contribution to book "Computer Algebra in Quantum Field Theory: Integration, 
Summation and Special Functions"}

\maketitle 

\abstract{We review the large $N$ method of calculating high order information 
on the renormalization group functions in a quantum field theory which is based
on conformal integration methods. As an example these techniques are applied to
a typical graph contributing to the $\beta$-function of $O(N)$ $\phi^4$ theory
at $O(1/N^2)$. The possible future directions for the large $N$ methods are 
discussed in light of the development of more recent techniques such as the 
Laporta algorithm.}

\vspace{-9.5cm}
\hspace{10cm}
{\bf LTH 970}

\vspace{9.5cm}
\section{Introduction}

One of the main problems in renormalization theory is the construction of the
renormalization group functions. These govern how the parameters of a quantum
field theory, such as the coupling constant, depend on scale. In situations
where one has to compare with precision data, this ordinarily requires
knowing the renormalization group functions to very high orders in a
perturbative expansion. The quantum field theories we have in mind are not only
the gauge theories of particle physics but also the scalar and fermionic ones 
which arise in condensed matter problems. These are central in understanding
phase transitions. To attain such precision in perturbative expansions means
that large numbers of Feynman diagrams have to be determined with the number of
graphs increasing with the loop order. Moreover, as the order increases the 
underlying integrals require more sophisticated methods in order to deduce 
their value analytically. The widely established methods of computing Feynman 
graphs will be reported elsewhere in this volume. Here we review an alternative
approach which complements explicit perturbative techniques. It does so in
such a way that for low loop orders there is overlap but at orders beyond
that already known part of the perturbative series can be deduced at {\em all}
orders within a certain approximation. This is known as the large $N$ or large
$\Nf$ method where $N$ is a parameter deriving from a symmetry of the theory 
such as a Lie group or the number of massless quark flavours, $\Nf$, in Quantum
Chromodynamics (QCD). In this method the Feynman graphs are related to those of
perturbation theory but because of the nature of the expansion parameter, the 
powers of the propagators appearing in such graphs are not the canonical value 
of unity but instead differ from unity by $O(\epsilon)$ where $\epsilon$ 
corresponds to the regularizing parameter of dimensional regularization. In 
addition beyond leading order in the $1/N$ expansion, the propagator powers
will include the anomalous dimensions in addition to the leading or canonical
dimension. Therefore, standard perturbative techniques such as integration by 
parts requires care in its use since one may not be able to actually reduce a 
graph to a simpler topology. Instead a different technique has had to be 
refined and developed. It is based on a conformal\index{conformal} property of 
Feynman integrals and we review it here in the context of the large $N$ 
methods. Though it has had some applications in perturbative computations.  

The article is organized as follows. We devote the next section to the 
notation and techniques of computing Feynman graphs using conformal methods in
$d$-dimensions. We focus on the general two loop self energy graph in the
subsequent section and review the work of \cite{Gracey:1,Gracey:2}, upon which 
this review is mostly based, and others in the methods of evaluating it. These 
techniques are then applied to a problem in scalar quantum field theory in
Section $4$ where a graph with $10$ internal integrations is evaluated 
{\em exactly} in $d$-dimensions. We conclude in Section $5$ with thoughts on
the direction in which the technique could be developed next given recent 
advances in the computation of Feynman graphs using conventional perturbative 
techniques. 

\section{Notation and Elementary Techniques}

We begin by introducing the notation we will use which will be based on 
\cite{Gracey:1,Gracey:2}. There Feynman graphs were represented in coordinate 
or configuration space notation. By this we mean that in writing a Feynman 
integral graphically the integration variables are represented as the vertices.
By contrast in momentum space representation the integration variables 
correspond to the momenta circulating around a loop. So in coordinate space
representation propagators are denoted by lines between two fixed points, as 
illustrated in Figure $1$.
\begin{center}
\begin{picture}(100,55)(0,-10)
\Line(0,20)(60,20)
\Vertex(0,20){1}
\Vertex(60,20){1}
\Text(0,10)[]{$x$}
\Text(30,30)[]{$\alpha$}
\Text(60,10)[]{$y$}
\Text(100,20)[]{~~ $\equiv ~~~ \frac{1}{((x-y)^2)^\alpha}$}
\end{picture}

{Figure $1$. Coordinate space propagator.}
\end{center}
There the power of the propagator is denoted by a number or symbol beside the
line. One can map between coordinate and momentum space representation by
using a Fourier transform. In the notation of \cite{Gracey:1,Gracey:2} we have
\begin{equation}
\frac{1}{(x^2)^\alpha} ~=~ \frac{a(\alpha)}{2^{2\alpha}\pi^\mu} \int_k
d^d k \, \frac{e^{ikx}}{(k^2)^{\mu-\alpha}}
\end{equation}
where $x$ is in coordinate space and $k$ is the conjugate momentum. Also for
shorthand we set
\begin{equation}
d ~=~ 2 \mu
\end{equation}
which is used throughout to avoid the appearance of $d/2$ in the Euler
$\Gamma$-function. This symbol should not be confused with the mass scale
appearing in renormalization group equations. Clearly 
\begin{equation}
a(\alpha) ~=~ \frac{\Gamma(\mu-\alpha)}{\Gamma(\alpha)}
\end{equation}
which is singular when $\alpha$~$=$~$\mu$~$+$~$n$ where $n$ is zero or a 
positive integer. Also $a(\alpha)$ vanishes at the negative integers. The
elementary identity
\begin{equation}
a(\alpha) a(\mu-\alpha) ~=~ 1
\end{equation}
follows trivially as does
\begin{equation}
a(\alpha) ~=~ \frac{a(\alpha-1)}{(\alpha-1)(\mu-\alpha)}
\end{equation}
from the $\Gamma$-function identity $\Gamma(z+1)$~$=$~$z\Gamma(z)$. With this 
notation the elementary one loop self energy graph in momentum space is 
replaced by chain integration in coordinate space representation. This is 
represented graphically in Figure $2$
\begin{center}
\begin{picture}(240,60)(0,-10)
\Line(0,20)(80,20)
\Vertex(0,20){1}
\Vertex(40,20){1}
\Vertex(80,20){1}
\Text(0,10)[]{$0$}
\Text(20,30)[]{$\alpha$}
\Text(60,30)[]{$\beta$}
\Text(40,10)[]{$y$}
\Text(80,10)[]{$x$}
\Text(140,20)[]{~~ $\equiv ~~~ \nu(\alpha,\beta,2\mu-\alpha-\beta)$} 
\Line(210,20)(260,20)
\Text(235,30)[]{$\alpha+\beta-\mu$}
\Text(210,10)[]{$0$}
\Text(260,10)[]{$x$}
\Vertex(210,20){1}
\Vertex(260,20){1}
\end{picture}

{Figure $2$. Chain integration.}
\end{center}
where, \cite{Gracey:1,Gracey:2}, 
\begin{equation}
\nu(\alpha,\beta,\gamma) ~=~ \pi^\mu a(\alpha) a(\beta) a(\gamma) ~.
\end{equation}
However, in practice Feynman graphs have more complicated integration points.
In other words in coordinate space representation one has more than two lines
intersecting at a point. Therefore, more involved integration techniques are
required to evaluate the Feynman integrals. One very useful technique is that
of uniqueness or conformal\index{conformal} integration which was introduced in
three dimensions in \cite{Gracey:3}. It has been developed in several ways 
subsequently and specifically to $d$-dimensions. For example, see 
\cite{Gracey:4}. We follow \cite{Gracey:1,Gracey:2} and use the rule 
represented in Figure $3$, where $z$ is the integration variable,
\begin{center}
\begin{picture}(240,100)(0,-10)
\Line(0,20)(20,40)
\Line(40,20)(20,40)
\Line(20,60)(20,40)
\Vertex(0,20){1}
\Vertex(20,40){1}
\Vertex(40,20){1}
\Vertex(20,60){1}
\Text(0,10)[]{$x$}
\Text(40,10)[]{$y$}
\Text(20,30)[]{$z$}
\Text(20,70)[]{$0$}
\Text(26,50)[]{$\alpha$}
\Text(7,35)[]{$\gamma$}
\Text(35,35)[]{$\beta$}
\Text(90,40)[]{~~ $\equiv ~~~ \nu(\alpha,\beta,\gamma)$} 
\Line(170,20)(210,20)
\Line(170,20)(190,60)
\Line(210,20)(190,60)
\Vertex(170,20){1}
\Vertex(190,60){1}
\Vertex(210,20){1}
\Text(162,15)[]{$x$}
\Text(218,15)[]{$y$}
\Text(190,70)[]{$0$}
\Text(190,10)[]{$\mu-\alpha$}
\Text(220,40)[]{$\mu-\gamma$}
\Text(160,40)[]{$\mu-\beta$}
\end{picture}

{Figure $3$. Conformal integration when 
$\alpha$~$+$~$\beta$~$+$~$\gamma$~$=$~$2\mu$.}
\end{center}
which follows when the sum of the exponents of the lines intersecting at the
$3$-point vertex add to the spacetime dimension
\begin{equation}
\alpha ~+~ \beta ~+~ \gamma ~=~ 2 \mu ~.
\label{Gracey:uniq}
\end{equation} 
This is known as the uniqueness condition. By the same token if a graph 
contains a triangle where the lines comprising the triangle sum to $\mu$ such 
as
\begin{equation}
(\mu - \alpha) ~+~ (\mu - \beta) ~+~ (\mu - \gamma) ~=~ \mu 
\end{equation}
as is the case in Figure $3$, then the unique triangle can be replaced by the 
vertex on the left side. There are several methods to establish the uniqueness 
integration rule. If one uses standard text book methods such as Feynman 
parameters then the integral over $z$ can be written as
\begin{eqnarray}
&& \frac{\pi^\mu \Gamma(\mu-\alpha) \Gamma(\alpha+\beta+\gamma-\mu)}
{\Gamma(\beta) \Gamma(\gamma) \Gamma(\mu)} \nonumber \\
&& \times \int_0^1 db 
\frac{b^{\beta-1}(1-b)^{\gamma-1}}{[b(1-b)(x-y)^2]^{\alpha+\beta+\gamma-\mu}}
\nonumber \\
&& \times {}_2F_1 \left( \alpha+\beta+\gamma-\mu,\alpha;\mu; 
-~ \frac{[bx+(1-b)y]}{b(1-b)(x-y)^2} \right)  
\end{eqnarray}
prior to using, (\ref{Gracey:uniq}). When that condition is set then the 
hypergeometric function\index{hypergeometric function} collapses to the 
geometric series and allows the integration over the Feynman parameter $b$ to 
proceed which results in
\begin{eqnarray}
&& \frac{\pi^\mu \Gamma(\mu-\alpha) \Gamma(\mu-\beta) \Gamma(\mu-\gamma)}
{\Gamma(2\mu-\beta-\gamma) \Gamma(\beta) \Gamma(\gamma) (y^2)^\alpha
[(x-y)^2]^{\beta+\gamma-\mu} } \nonumber \\
&& \times {}_2 F_1 \left( \alpha, \mu-\gamma; 2\mu - \beta - \gamma ;
1 - \frac{x^2}{y^2} \right)
\end{eqnarray}
Applying the uniqueness condition a second time produces the right hand side of
Figure $3$ since the hypergeometric function\index{hypergeometric function} 
again reduces to the geometric series. This is in such a way that the canonical
propagators emerge. 

An alternative method is to apply a 
conformal\index{conformal} transformation on the coordinates of the integral, 
\cite{Gracey:2}. In this approach, which is applicable to any graph in general,
one external point is labelled as an origin and given $0$ as a coordinate. The 
other points are denoted by coordinates $x$, $y$ and $z$. The conformal 
transformation changes the integration coordinate as well as the external 
points through
\begin{equation}
x_\mu ~ \rightarrow ~ \frac{x_\mu}{x^2} ~.
\label{Gracey:confr1}
\end{equation}
Thus for two coordinates $y$ and $z$ undergoing such a transformation we have
the lemma
\begin{equation}
(y-z)^2 ~ \rightarrow ~ \frac{(y-z)^2}{y^2 z^2} ~.
\label{Gracey:confr2}
\end{equation}
An integration measure also produces contributions to the lines joining to the 
origin since 
\begin{equation}
d^d z ~ \rightarrow ~ \frac{d^d z}{(z^2)^{2\mu}} ~.
\end{equation}
Therefore, for the vertex on the left side of Figure $3$ this transformation
produces the intermediate integral of Figure $4$. 
\begin{center}
\begin{picture}(88,120)(-20,-10)
\Line(-10,20)(20,40)
\Line(50,20)(20,40)
\Line(20,80)(20,40)
\Line(-10,20)(20,80)
\Line(50,20)(20,80)
\Vertex(-10,20){1}
\Vertex(20,40){1}
\Vertex(50,20){1}
\Vertex(20,80){1}
\Text(-10,10)[]{$x$}
\Text(50,10)[]{$y$}
\Text(20,30)[]{$z$}
\Text(20,90)[]{$0$}
\Text(24,50)[]{$\bar{\alpha}$}
\Text(8,22)[]{$\gamma$}
\Text(32,22)[]{$\beta$}
\Text(-14,53)[]{$2\mu-\gamma$}
\Text(54,53)[]{$2\mu-\beta$}
\end{picture}

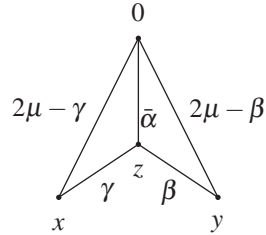
{Figure $4$. Vertex of Figure $3$ after a conformal transformation with base at
$0$ where $\bar{\alpha}$~$=$~$2\mu$~$-$~$\alpha$~$-$~$\beta$~$-$~$\gamma$.}
\end{center}
To complete the integration requires setting the uniqueness condition
(\ref{Gracey:uniq}) which produces a chain integral since the line from $0$ to
$z$ is absent from the graph. To complete the derivation one undoes the 
original conformal\index{conformal} transformations to produce the right hand 
side of Figure $3$. If one compares the two derivations, the latter is in fact 
of more practical use. This is because it avoids the use of writing the 
original integral in terms of Feynman parameters which would become tedious for
higher order cases. Also it is simple to implement graphically.

Having recalled the derivation of the uniqueness rule it is straightforward to 
see that there is a natural extension. In the first derivation there was not a
unique way to collapse the hypergeometric function to an elementary type of
propagator. Instead this will happen if the sum of the exponents is
$(2\mu$~$+$~$n)$ where $n$ is a positive integer. Athough the collapse in this
case will not be to the geometric series, it will reduce to simple algebraic
functions which are of the propagator type. So, for instance, when 
$n$~$=$~$1$ we have the result of Figure $5$, \cite{Gracey:5},
\begin{center}
\begin{picture}(240,240)(0,0)
\Line(0,180)(20,200)
\Line(40,180)(20,200)
\Line(20,220)(20,200)
\Vertex(0,180){1}
\Vertex(20,200){1}
\Vertex(40,180){1}
\Vertex(20,200){1}
\Text(0,170)[]{$x$}
\Text(40,170)[]{$y$}
\Text(20,190)[]{$z$}
\Text(20,230)[]{$0$}
\Text(26,210)[]{$\alpha$}
\Text(7,195)[]{$\gamma$}
\Text(35,195)[]{$\beta$}
\Text(90,200)[]{~~ $\equiv ~~~ 
\frac{\nu(\alpha-1,\beta-1,\gamma)}{(\alpha-1)(\beta-1)}$} 
\Line(170,180)(210,180)
\Line(170,180)(190,220)
\Line(210,180)(190,220)
\Vertex(170,180){1}
\Vertex(190,220){1}
\Vertex(210,180){1}
\Text(162,175)[]{$x$}
\Text(218,175)[]{$y$}
\Text(190,230)[]{$0$}
\Text(190,170)[]{$\mu-\alpha+1$}
\Text(220,200)[]{$\mu-\gamma$}
\Text(156,200)[]{$\mu-\beta+1$}
\Text(90,120)[]{~~ $~ + ~~ 
\frac{\nu(\alpha-1,\beta,\gamma-1)}{(\alpha-1)(\gamma-1)}$} 
\Line(170,100)(210,100)
\Line(170,100)(190,140)
\Line(210,100)(190,140)
\Vertex(170,100){1}
\Vertex(190,140){1}
\Vertex(210,100){1}
\Text(162,95)[]{$x$}
\Text(218,95)[]{$y$}
\Text(190,150)[]{$0$}
\Text(190,90)[]{$\mu-\alpha+1$}
\Text(228,120)[]{$\mu-\gamma+1$}
\Text(160,120)[]{$\mu-\beta$}
\Text(90,40)[]{~~ $~ + ~~ 
\frac{\nu(\alpha,\beta-1,\gamma-1)}{(\beta-1)(\gamma-1)}$} 
\Line(170,20)(210,20)
\Line(170,20)(190,60)
\Line(210,20)(190,60)
\Vertex(170,20){1}
\Vertex(190,60){1}
\Vertex(210,20){1}
\Text(162,15)[]{$x$}
\Text(218,15)[]{$y$}
\Text(190,70)[]{$0$}
\Text(190,10)[]{$\mu-\alpha$}
\Text(228,40)[]{$\mu-\gamma+1$}
\Text(156,40)[]{$\mu-\beta+1$}
\end{picture}

{Figure $5$. Conformal integration when 
$\alpha$~$+$~$\beta$~$+$~$\gamma$~$=$~$2\mu$~$+$~$1$.}
\end{center}
A similar rule has been constructed and used in \cite{Gracey:4}. We will use
Figure $4$ later in order to simplify various integrals.

\section{Two Loop Self Energy Graph}

We can illustrate some of the techniques of conformal\index{conformal}
integration by considering the massless two loop self energy graph with 
arbitrary powers, $\alpha_i$ on the propagators. It is illustrated in Figure
$6$
\begin{center}
\begin{picture}(100,90)(-10,0)
\Line(0,50)(40,25)
\Line(0,50)(40,75)
\Line(80,50)(40,25)
\Line(80,50)(40,75)
\Line(40,25)(40,75)
\Vertex(0,50){1.2}
\Vertex(40,25){1.2}
\Vertex(40,75){1.2}
\Vertex(80,50){1.2}
\Text(-7,50)[]{$0$}
\Text(87,50)[]{$x$}
\Text(40,17)[]{$z$}
\Text(40,83)[]{$y$}
\Text(10,33)[]{$\alpha_4$}
\Text(10,67)[]{$\alpha_1$}
\Text(70,33)[]{$\alpha_3$}
\Text(70,67)[]{$\alpha_2$}
\Text(50,50)[]{$\alpha_5$}
\end{picture}

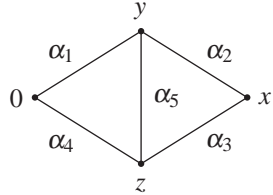
{Figure $6$. Two loop self energy graph in coordinate space representation.}
\end{center}
where we have used the coordinate space representation. Thus the vertices are 
integrated over rather than the loop momenta. To clarify, the integral of 
Figure $6$ is
\begin{equation}
I(\alpha_1,\alpha_2,\alpha_3,\alpha_4,\alpha_5) ~=~ \int_{yz}
\frac{1}{(y^2)^{\alpha_1} ((x-y)^2)^{\alpha_2} ((x-z)^2)^{\alpha_3} 
(z^2)^{\alpha_4} ((y-z)^2)^{\alpha_5}} 
\end{equation}
where $\int_y$~$=$~$\int \frac{d^dy}{(2\pi)^d}$. The structure of this
integral has been widely studied and we briefly highlight several properties of
relevance. The analysis of \cite{Gracey:6,Gracey:7} determined that the 
symmetry group of the graph was $Z_2$~$\times$~$S_6$ which has $1440$ elements.
Exploiting this the $\epsilon$ expansion of the integral in 
$d$~$=$~$4$~$-$~$2\epsilon$ with propagator powers of order $\epsilon$ from 
unity was determined up to $O(\epsilon^6)$, \cite{Gracey:6,Gracey:7}. At 
$O(\epsilon^5)$ it was discovered that the first multi-zeta
value\index{mutiple zeta} occurred, \cite{Gracey:7}. Specifically
\begin{eqnarray}
I(1,1,\mu-1,1,\mu-1) &=& 6 \zeta_3 + 9 \zeta_4 \epsilon + 7 \zeta_5 \epsilon^5
\nonumber \\
&& +~ \frac{5}{2} \left[ \zeta_6 - 2 \zeta_3^2 \right] \epsilon^3
- \frac{1}{8} \left[ 91 \zeta_7 + 120 \zeta_3 \zeta_4 \right] \epsilon^4
\nonumber \\
&& +~ \frac{1}{81920} \left[ 653440 \zeta_5 \zeta_3 - 7059417 \zeta_8 
+ 576 F_{53} \right] \epsilon^5 \nonumber \\
&& +~ O(\epsilon^6)
\label{Gracey:intdjb}
\end{eqnarray}
where $\zeta_z$ is the Riemann zeta function and $F_{53}$~$=$~$\sum_{n>m>0} 
\frac{1}{n^5m^3}$ in the original notation of \cite{Gracey:8}. Subsequent to 
this it has been shown that the only numbers which appear in the full series 
expansion in $\epsilon$ are mutiple zeta values, \cite{Gracey:9}. While the 
work of \cite{Gracey:6,Gracey:7} illustrated the power of group theory to 
evaluate master integrals explicitly, using conformal\index{conformal}
integration allows one to relate two loop self energy integrals by exploiting 
the masslessness of the original diagrams. This was originally developed in 
\cite{Gracey:1,Gracey:2} and we summarize that here as there appears to be 
scope nowadays to take this method to three and higher loop order graphs.

The transformations developed in \cite{Gracey:2} fall into several classes. The
first is that derived from the elementary use of the Fourier transform. Writing
\begin{equation}
I(\alpha_1,\alpha_2,\alpha_3,\alpha_4,\alpha_5) ~=~ 
\frac{\Gamma}{(x^2)^{D-2\mu}}
\label{Gracey:idef}
\end{equation}
where $D$~$=$~$\sum_{i=1}^5 \alpha_i$ and $\Gamma$ is independent of $x$ and
corresponds to the value of the integral, then taking the Fourier transform 
produces an integral which is also the two loop self energy. Though the 
propagator powers are different. In this sense one can say that the graph is 
self-dual which is not a property all Feynman graphs have. Thus, 
\cite{Gracey:2},
\begin{equation}
I(\alpha_1,\alpha_2,\alpha_3,\alpha_4,\alpha_5) ~=~ 
\frac{\prod_{i=1}^5 a(\alpha_i)}{a(D-2\mu)}
I(\mu-\alpha_2,\mu-\alpha_3,\mu-\alpha_4,\mu-\alpha_1,\mu-\alpha_5) ~.
\end{equation}
This transformation is known as the momentum representation or MR. It can be 
easily generalized to other topologies and there is a simple graphical rule for
this. Although not immediately apparent from the self energy because of the 
self-duality, each $3$-valent vertex of the original graph has an associated 
triangle in the dual graph. For other topologies $4$-valent vertices are mapped
to squares and $5$-valent vertices to pentagons with a clear generalization 
pattern. 

A set of less obvious transformations can be deduced from the uniqueness 
condition. First, we define the shorthand notation, \cite{Gracey:2},
\begin{eqnarray}
s_1 &=& \alpha_1 + \alpha_2 + \alpha_5 ~~~,~~~ 
s_2 ~=~ \alpha_3 + \alpha_4 + \alpha_5 \nonumber \\
t_1 &=& \alpha_1 + \alpha_4 + \alpha_5 ~~~,~~~
t_2 ~=~ \alpha_2 + \alpha_3 + \alpha_5 
\end{eqnarray}
and illustrate the technique for one case. If one considers the central
propagator it can be replaced by a chain integral. Although there are an
infinite number of ways of doing this one can choose the exponents of the chain
so that the top vertex is unique. In other words
\begin{eqnarray}
&& \frac{1}{((y-z)^2)^{\alpha_5}} \nonumber \\ 
&& =~ \frac{1}{\nu(2\mu-\alpha_1-\alpha_2,s_1-\mu,\mu-\alpha_5)} \int_u 
\frac{1}{((y-u)^2)^{2\mu-\alpha_1-\alpha_2} ((u-z)^2)^{s_1-\mu}} 
\end{eqnarray}
where $u$ is the intermediate integration point. As the $y$ vertex of Figure 
$6$ is now unique the conformal integration rule can be used to rewrite the 
integral. This results in, \cite{Gracey:2}, 
\begin{equation}
I(\alpha_1,\alpha_2,\alpha_3,\alpha_4,\alpha_5) ~=~ 
\frac{a(\alpha_1) a(\alpha_2) a(\alpha_5)}{a(s_1-\mu)} 
I(\mu-\alpha_2,\mu-\alpha_1,\alpha_3,\alpha_4,s_1-\mu) ~. 
\end{equation}
In the notation of \cite{Gracey:2} this transformation is known as $\uparrow$. 
It is elementary to see that there are five other such transformations which 
are denoted by $\nearrow$, $\nwarrow$, $\downarrow$, $\searrow$ and $\swarrow$.
The syntax is that when an arrow points in a general upwards direction it is a 
transformation on the $y$ vertex and by contrast in a downwards direction it 
relates to the $z$ vertex. The propagator which one replaces by a chain to make
the vertex unique is in correspondence with the direction of the arrow. While 
these six transformations operate on the internal vertices there are two which 
act on each of the external vertices. One can complete the uniqueness of one of
these by realizing that the integral itself is a propagator with power 
$(D-2\mu)$ as indicated in (\ref{Gracey:idef}), \cite{Gracey:2}. For example, 
if the right external point is chosen as the base integration vertex then the 
appending propagator has power $(2\mu-\alpha_2-\alpha_3)$. This produces 
\begin{equation}
I(\alpha_1,\alpha_2,\alpha_3,\alpha_4,\alpha_5) ~=~ 
\frac{a(\alpha_2) a(\alpha_3)}{a(D-2\mu)a(2\mu-t_1)} 
I(\alpha_1,\mu-\alpha_3,\mu-\alpha_2,\alpha_4,t_2-\mu)  
\end{equation}
and this is denoted by $\leftarrow$. The corresponding transformation on the 
left external point is called $\rightarrow$. 

The final set of transformations are based on the conformal\index{conformal}
transformations (\ref{Gracey:confr1}) and (\ref{Gracey:confr2}) together with 
the effect they have on the two vertex measures, \cite{Gracey:2}. One can 
choose either of the external vertices as the origin of the transformation. 
Once decided the result of the conformal transformation is that all propagators
joining to the origin have their powers changed to the difference of $2\mu$ and
the sum of the exponents at the point at the other end of that propagator. This
means all points including those not directly connected to the base point in 
the first place. For the two loop self energy there are no such points but for 
higher loop graphs this will be the case. We will give an example of this in
Section $4$. Thus the conformal left transformation is, \cite{Gracey:2},  
\begin{equation}
I(\alpha_1,\alpha_2,\alpha_3,\alpha_4,\alpha_5) ~=~ 
I(2\mu-s_1,\alpha_2,\alpha_3,2\mu-s_2,\alpha_5)  
\end{equation}
where there is no $\Gamma$-function factor and this is denoted by CL in 
contrast to CR which is the transformation based on the right external vertex 
as the origin of the conformal transformation. The full set of transformations
and the result of applying each to the graph of Figure $6$ are summarized in a
Table in \cite{Gracey:2}. However, as brief examples of the transformations the
integral of (\ref{Gracey:intdjb}) is related as follows
\begin{eqnarray} 
I(1,1,\mu-1,1,\mu-1) & \stackrel{\uparrow}{=} & I(\mu-1,\mu-1,1,\mu-1,1) 
\nonumber \\
& \stackrel{\mbox{\footnotesize{CR}}}{=} & I(\mu-1,1,1,1,\mu-1) ~.
\end{eqnarray}
Though the latter follows from a simple rotation of the integral as well. 

Aside from the transformations there are other techniques which allow one to
evaluate the two loop self energy and higher order graphs. Perhaps the most
exploited is that of integration by parts which was introduced for 
(\ref{Gracey:idef}) in \cite{Gracey:10}. It determined that the first term in 
the $\epsilon$ expansion of $I(1,1,1,1,1)$ was $6 \zeta_3$ and has also been 
used in other applications, \cite{Gracey:2}. Indeed more recently the technique
has been developed by Laporta\index{Laporta} in \cite{Gracey:11} to produce an 
algorithm which relates all integrals in a Feynman graph to a base set of 
master integrals. These can then be evaluated by direct methods to complete the
overall computation. In the coordinate space representation we use here the 
basic rule is given in Figure $7$
\begin{center}
\begin{picture}(290,160)(0,0)
\Line(0,90)(20,110)
\Line(40,90)(20,110)
\Line(20,130)(20,110)
\Vertex(0,90){1}
\Vertex(20,110){1}
\Vertex(40,90){1}
\Vertex(20,110){1}
\Text(0,80)[]{$x$}
\Text(40,80)[]{$y$}
\Text(20,100)[]{$z$}
\Text(20,140)[]{$0$}
\Text(26,120)[]{$\alpha$}
\Text(7,105)[]{$\gamma$}
\Text(35,105)[]{$\beta$}
\Text(90,110)[]{~~ $\equiv ~~
\frac{\beta}{(2\mu-2\alpha-\beta-\gamma)} \biggl( \frac{}{} \biggr.$} 
\Line(140,90)(160,110)
\Line(180,90)(160,110)
\Line(160,130)(160,110)
\Vertex(140,90){1}
\Vertex(180,90){1}
\Vertex(160,130){1}
\Vertex(160,110){1}
\Text(167,120)[]{$-$}
\Text(175,104)[]{$+$}
\Text(200,110)[]{~~ $- ~~~ $} 
\Line(210,90)(230,110)
\Line(250,90)(230,110)
\Line(230,130)(230,110)
\Line(250,90)(230,130)
\Vertex(210,90){1}
\Vertex(250,90){1}
\Vertex(230,130){1}
\Vertex(230,130){1}
\Text(247,117)[]{$-$}
\Text(236,96)[]{$+$}
\Text(260,110)[]{$\biggl. \frac{}{} \biggr)$} 
\Text(90,40)[]{~~ $~~ + ~ 
\frac{\gamma}{(2\mu-2\alpha-\beta-\gamma)} \biggl( \frac{}{} \biggr.$} 
\Line(140,20)(160,40)
\Line(180,20)(160,40)
\Line(160,60)(160,40)
\Vertex(140,20){1}
\Vertex(180,20){1}
\Vertex(160,60){1}
\Vertex(160,40){1}
\Text(167,50)[]{$-$}
\Text(145,34)[]{$+$}
\Text(200,40)[]{~~ $- ~~~ $} 
\Line(210,20)(230,40)
\Line(250,20)(230,40)
\Line(230,60)(230,40)
\Line(210,20)(230,60)
\Vertex(210,20){1}
\Vertex(250,20){1}
\Vertex(230,60){1}
\Vertex(230,60){1}
\Text(217,47)[]{$-$}
\Text(226,26)[]{$+$}
\Text(260,40)[]{$\biggl. \frac{}{} \biggr) $} 
\end{picture}

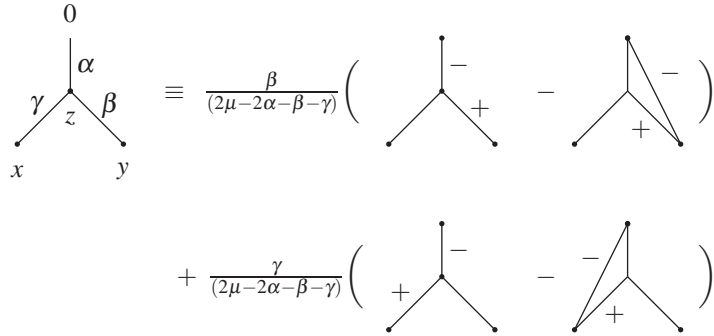
{Figure $7$. Integration by parts in coordinate space representation.}
\end{center}
where the $+$ or $-$ on a line indicates that the power of that propagator is
increased or decreased by unity. For example, with this, \cite{Gracey:10},
\begin{equation}
I(1,1,1,1,1) ~=~ \frac{\nu(1,1,2\mu-2)}{(\mu-2)} \left[ \nu(1,2,2\mu-3) -
\nu(3-\mu,2,3\mu-5) \right]
\end{equation}
which can be expanded in powers of $\epsilon$. Clearly the series can only
involve rationals and $\zeta_n$. Indeed the rule can also be applied to more
general cases. In \cite{Gracey:2} it was shown that
\begin{eqnarray}
I(\alpha,\mu-1,\mu-1,\beta,\mu-1) &=& \frac{a(2\mu-2)}{\Gamma(\mu-1)} \left[
\frac{a(\alpha)a(2-\alpha)}{(1-\beta)(\alpha+\beta-2)} \right. \nonumber \\
&& \left. ~~~~~~~~~~~~~~~~~~~+~ 
\frac{a(\alpha+\beta-1)a(3-\alpha-\beta)}{(\alpha-1)(\beta-1)} \right.
\nonumber \\ 
&& \left. ~~~~~~~~~~~~~~~~~~~+~ 
\frac{a(\beta)a(2-\beta)}{(1-\alpha)(\alpha+\beta-2)} 
\right]
\end{eqnarray}
for arbitrary $\alpha$ and $\beta$. However, not all graphs can be integrated
by parts. An example of such a case is $I(1,\alpha,\beta,\gamma,1)$ for 
non-unit $\alpha$, $\beta$ and $\gamma$. Another example is 
(\ref{Gracey:intdjb}), \cite{Gracey:7}, whose expansion has a non-Riemann zeta 
value at some point in the expansion. Indeed this is perhaps an indication of 
an obstruction to integrability.

While integration by parts allows one to reduce the powers of various
propagators 
\begin{center}
\begin{picture}(300,230)(-10,0)
\Line(10,190)(50,165)
\Line(10,190)(50,215)
\Line(90,190)(50,165)
\Line(90,190)(50,215)
\Line(50,165)(50,215)
\Vertex(10,190){1.2}
\Vertex(50,165){1.2}
\Vertex(50,215){1.2}
\Vertex(90,190){1.2}
\Text(3,190)[]{$0$}
\Text(97,190)[]{$x$}
\Text(50,157)[]{$z$}
\Text(50,223)[]{$y$}
\Text(20,173)[]{$\alpha_4$}
\Text(20,207)[]{$\alpha_1$}
\Text(80,173)[]{$\alpha_3$}
\Text(80,207)[]{$\alpha_2$}
\Text(60,190)[]{$\alpha_5$}
\Text(150,190)[]{~~ $= ~~~ 
\frac{\alpha_1(\mu-\alpha_1-1)}{(\alpha_2-1)(\alpha_5-1)}$} 
\Line(200,190)(240,165)
\Line(200,190)(240,215)
\Line(280,190)(240,165)
\Line(280,190)(240,215)
\Line(240,165)(240,215)
\Vertex(200,190){1.2}
\Vertex(240,165){1.2}
\Vertex(240,215){1.2}
\Vertex(280,190){1.2}
\Text(210,207)[]{$+$}
\Text(270,173)[]{$+$}
\Text(270,207)[]{$-$}
\Text(250,190)[]{$-$}
\Text(150,120)[]{~~ $+ ~~~ 
\frac{(\alpha_2+\alpha_5-\mu-1)}{(\alpha_2-1)}$} 
\Line(200,120)(240,95)
\Line(200,120)(240,145)
\Line(280,120)(240,95)
\Line(280,120)(240,145)
\Line(240,95)(240,145)
\Vertex(200,120){1.2}
\Vertex(240,95){1.2}
\Vertex(240,145){1.2}
\Vertex(280,120){1.2}
\Text(270,103)[]{$+$}
\Text(270,137)[]{$-$}
\Text(150,50)[]{~~ $+ ~~~ 
\frac{(\alpha_2+\alpha_5-\mu-1)}{(\alpha_5-1)}$} 
\Line(200,50)(240,25)
\Line(200,50)(240,75)
\Line(280,50)(240,25)
\Line(280,50)(240,75)
\Line(240,25)(240,75)
\Vertex(200,50){1.2}
\Vertex(240,25){1.2}
\Vertex(240,75){1.2}
\Vertex(280,50){1.2}
\Text(270,33)[]{$+$}
\Text(250,50)[]{$-$}
\end{picture}

{Figure $8$. Reduction formula for two loop self energy based on the
generalized $\nearrow$ transformation.}
\end{center}
by unity within a Feynman diagram it is not the only method to 
achieve this. A modification of the uniqueness method can be used to derive
rules similar to Figure $7$. Specifically if one chooses the exponents of the
propagators to be $(2\mu+1)$ then one finds the extension given in Figure $5$. 
Using this rule and repeating the analysis of the transformations on the two 
loop self energy graph provides relations specific to this topology,
\cite{Gracey:5}. For instance, extending $\nearrow$ to have the upper vertex 
exponents summing to $(2\mu+1)$ gives the relation in Figure $8$ where the $+$ 
or $-$ on the right side indicates that the exponent of that line is increased 
or decreased by unity.  
In Figure $8$ provided $\alpha_2$~$\neq$~$1$ and $\alpha_5$~$\neq$~$1$ then the
powers of the respective propagators can be reduced by unity. However, this 
restriction is a drawback if one wishes to reduce graphs which have unit
exponents. Instead it is possible to extend the method which produced the
relation of Figure $8$. For instance, rather than begin with the general two 
loop self energy and applying the generalized uniqueness rule, one can use one 
of the transformations of \cite{Gracey:2} and then apply a rule like that of 
Figure $8$ before applying the transformation inverse to the original one. In 
this way one can build up a suite of relations. 
\begin{center}
\begin{picture}(300,230)(-10,0)
\Line(10,190)(50,165)
\Line(10,190)(50,215)
\Line(90,190)(50,165)
\Line(90,190)(50,215)
\Line(50,165)(50,215)
\Vertex(10,190){1.2}
\Vertex(50,165){1.2}
\Vertex(50,215){1.2}
\Vertex(90,190){1.2}
\Text(3,190)[]{$0$}
\Text(97,190)[]{$x$}
\Text(50,157)[]{$z$}
\Text(50,223)[]{$y$}
\Text(20,173)[]{$\alpha_4$}
\Text(20,207)[]{$\alpha_1$}
\Text(80,173)[]{$\alpha_3$}
\Text(80,207)[]{$\alpha_2$}
\Text(60,190)[]{$\alpha_5$}
\Text(150,190)[]{~~ $= ~~~ 
\frac{(2\mu-s_1)(2\mu-s_2)}{(2\mu-t_2)(t_2-\mu-1)}$} 
\Line(200,190)(240,165)
\Line(200,190)(240,215)
\Line(280,190)(240,165)
\Line(280,190)(240,215)
\Line(240,165)(240,215)
\Vertex(200,190){1.2}
\Vertex(240,165){1.2}
\Vertex(240,215){1.2}
\Vertex(280,190){1.2}
\Text(250,190)[]{$-$}
\Text(160,120)[]{~~ $+ ~~~ 
\frac{(2\mu-s_2)(D+\alpha_5-3\mu-1)}{(2\mu-t_2)(t_2-\mu-1)}$}
\Line(220,120)(260,95)
\Line(220,120)(260,145)
\Line(300,120)(260,95)
\Line(300,120)(260,145)
\Line(260,95)(260,145)
\Vertex(220,120){1.2}
\Vertex(260,95){1.2}
\Vertex(260,145){1.2}
\Vertex(300,120){1.2}
\Text(290,103)[]{$-$}
\Text(160,50)[]{~~ $+ ~~~ 
\frac{(2\mu-s_1)(D+\alpha_5-3\mu-1)}{(2\mu-t_2)(t_2-\mu-1)}$}
\Line(220,50)(260,25)
\Line(220,50)(260,75)
\Line(300,50)(260,25)
\Line(300,50)(260,75)
\Line(260,25)(260,75)
\Vertex(220,50){1.2}
\Vertex(260,25){1.2}
\Vertex(260,75){1.2}
\Vertex(300,50){1.2}
\Text(290,67)[]{$-$}
\end{picture}

{Figure $9$. Another reduction formula for two loop self energy.}
\end{center}

One such useful relation is illustrated in Figure $9$ which is derived in 
several stages. The first is to construct a relation similar to that of Figure
$9$ by first applying $\leftarrow$ to the graph of Figure $6$ and then undoing 
it by applying the rule of Figure $5$ to the same external vertex. This 
produces a relation where $t_2$ increases by unity in each of the three 
resulting graphs. The second stage is to apply this rule to the graph of Figure
$6$ after a CR transformation has been enacted. To complete the derivation the 
final step is to undo with another CR transformation. Thus the $t_2$ value of 
each graph on the right hand side of Figure $9$ is one less than that of the 
graph on the left side. This reduction has coefficients on the right hand side 
which are non-singular for unit propagators. Other rules can be derived by this
method and a fuller set are recorded in Appendix B of \cite{Gracey:12}. It is 
worth noting that similar rules based on the generalized uniqueness where 
developed in \cite{Gracey:4}.

\section{QFT Application}

Having discussed the general techniques for determining massless Feynman
integrals using conformal\index{conformal} methods, we illustrate their 
usefulness in a practical problem in a quantum field theory. Specifically we 
focus on the determination of the critical exponents at a phase transition in 
various models in the large $N$ expansion. The background which we describe 
here is based on a series of articles, \cite{Gracey:1,Gracey:2,Gracey:13},
where exponents were determined in $d$-dimensions at $O(1/N^2)$ and $O(1/N^3)$.
The fact that $d$-dimensional results are computable means that information on 
the renormalization group functions can be deduced in various spacetime 
dimensions. This is due to a special feature of critical point field theories 
and that is that at a non-trivial fixed point of the renormalization group flow
the critical exponents correspond to the associated renormalization group
function at that fixed point. Thus information on the renormalization group 
functions is encoded in these exponents. Moreover, at a fixed point several 
quantum field theories can lie in the same universality class despite having 
different structures. This is invariably as a consequence of a common 
interaction in the Lagrangian. Thus the same exponents can be used to access 
the structure of the renormalization group functions of two different theories.
Further, as the spacetime dimension $d$ is not used as a regulator, information
on the exponents can be deduced simultaneously in several different dimensions 
such as three and four. For more background to the use of the renormalization 
group equation at near criticality in quantum field theories see, for example, 
\cite{Gracey:14}.

For the application of the conformal\index{conformal} methods we consider here 
we concentrate on the $O(N)$ nonlinear $\sigma$ model which is critically 
equivalent in $d$-dimensions to $O(N)$ $\phi^4$ theory. For the latter theory 
the Lagrangian is
\begin{equation}
L ~=~ \frac{1}{2} ( \partial_\mu \phi^i )^2 ~+~ \frac{g}{8} (\phi^i \phi^i)^2
\end{equation}
where $g$ is the coupling constant and $1$~$\leq$~$i$~$\leq$~$N$. Introducing 
an auxiliary field $\sigma$ equates this Lagrangian to
\begin{equation}
L ~=~ \frac{1}{2} ( \partial_\mu \phi^i )^2 ~+~ \frac{1}{2} \sigma 
(\phi^i \phi^i) ~-~ \frac{\sigma^2}{2g} ~.
\end{equation}
At criticality it is the interaction which drives the dynamics and thus it is
straightforward to see that in this formulation the Lagrangian interaction is
the same as that of the $O(N)$ nonlinear $\sigma$ model when the fields are
constrained to lie on an $(N-1)$-dimensional sphere. The constraint would have
a final term linear in $\sigma$ rather than a quadratic one together with a 
different coupling constant. This essentially is the origin of both field
theories being in the same universality class. The linear or quadratic terms in
$\sigma$ at criticality serve effectively to define the structure of the 
propagators. In coordinate space representation these are, 
\cite{Gracey:1,Gracey:2},
\begin{equation}
\langle \phi^i(0) \phi^j(x) \rangle ~=~ 
\frac{\delta^{ij}A}{(x^2)^\alpha} ~~~~,~~~~ 
\langle \sigma(0) \sigma(x) \rangle ~=~ \frac{B}{(x^2)^\beta} 
\label{Gracey:propdef}
\end{equation}
where $A$ and $B$ are $x$-independent amplitudes and $\alpha$ and $\beta$ are
the scaling dimensions of the fields. The latter comprise two parts. The first 
is the canonical dimension and the other is the anomalous dimension. Here
\begin{equation}
\alpha ~=~ \mu ~-~ 1 ~+~ \half \eta ~~~~,~~~~ \beta ~=~ 2 ~-~ \eta ~-~ \chi
\label{Gracey:expdef}
\end{equation}
where $\eta$ is the anomalous dimension of $\phi^i$ and $\chi$ is the vertex
anomalous dimension. The former is related to the renormalization group
function which is also termed the anomalous dimension, $\gamma(g)$, by
\begin{equation}
\eta ~=~ \gamma(g_c)
\end{equation}
where $g_c$ is the value of the coupling constant at the critical point,
\begin{equation}
\gamma(g) ~=~ \mu \frac{d~}{d\mu} \ln Z_\phi
\label{Gracey:gamdef}
\end{equation}
and $Z_\phi$ is the wave function renormalization constant. (In 
(\ref{Gracey:gamdef}) we have temporarily used $\mu$ to denote the standard 
renormalization group scale that underlies any renormalization group equation.)
To determine the values of the exponents to a particular order in $1/N$ 
requires solving the skeleton Schwinger-Dyson equation for the $2$-point 
functions at the same order. We do not discuss that formalism here, which can 
be found in \cite{Gracey:1,Gracey:2}, as our focus is rather on the evaluation 
of the Feynman graphs contributing to these equations. Though we should say 
that the presence of the non-zero anomalous dimensions in the propagators means
that in $2$-point functions there are no self energy corrections on any 
internal propagator as otherwise there would be double counting. So the number 
of graphs to consider is smaller than the corresponding perturbative case.  

The coupling constant at the critical point is denoted by $g_c$ and is defined 
as a nontrivial zero of the $\beta$-function, $\beta(g_c)$~$=$~$0$. As we are 
working in $d$-dimensions such a non-trivial zero exists in our theories since 
away from the spacetime dimension where the theory is renormalizable the 
coupling constant becomes dimensionful. Hence the first term of the 
$d$-dimensional $\beta$-function depends on $d$. Moreover, $g_c$ will depend on
the parameters of the theory which in our case here is $N$. Thus 
$g_c$~$=$~$g_c(d,N)$. Similarly $\eta$~$=$~$\eta(d,N)$ and 
$\chi$~$=$~$\chi(d,N)$. These can all be expanded in powers of $1/N$ where $N$ 
is large in such a way that the coefficients of $1/N$ are $d$-dependent. Thence
if one expresses these coefficients in powers of $\epsilon$ where 
$d$~$=$~$4$~$-$~$2\epsilon$ for $\phi^4$ theory or 
$d$~$=$~$2$~$+$~$\bar{\epsilon}$ for the nonlinear $\sigma$ model, then one can 
deduce the coefficients in the corresponding renormalization group equation to 
{\em all} orders in perturbation theory at that order in $1/N$. In this respect
it is important to note that in the large $N$ expansion $\epsilon$ or
$\bar{\epsilon}$ do not play the role of a regulator as they would do in
conventional perturbation theory.

Instead to see the origin of where a regulator is required one should consider 
the simple two loop contribution to the $\sigma$ self energy graph given in 
Figure $10$ in coordinate space representation. 
\begin{center}
\begin{picture}(100,100)(-10,0)
\Line(0,50)(40,25)
\Line(0,50)(40,75)
\Line(80,50)(40,25)
\Line(80,50)(40,75)
\Line(40,25)(40,75)
\Vertex(0,50){1.2}
\Vertex(40,25){1.2}
\Vertex(40,75){1.2}
\Vertex(80,50){1.2}
\Text(-7,50)[]{$0$}
\Text(87,50)[]{$x$}
\Text(40,17)[]{$z$}
\Text(40,83)[]{$y$}
\Text(10,33)[]{$\alpha$}
\Text(10,67)[]{$\alpha$}
\Text(70,33)[]{$\alpha$}
\Text(70,67)[]{$\alpha$}
\Text(48,50)[]{$\beta$}
\end{picture}

{Figure $10$. Two loop self energy for $\sigma$.}
\end{center}
To use conformal methods one has to check the sum of the exponents at a vertex 
in coordinate space representation. From (\ref{Gracey:expdef}) one can see that
\begin{equation}
2 \alpha ~+~ \beta ~=~ 2\mu ~-~ \chi ~.
\end{equation}
However, from the structure of the renormalization group equation at
criticality the anomalous dimensions $\eta$ and $\chi$ begin as $O(1/N)$. More,
specifically
\begin{equation}
\eta ~=~ \sum_{i=1}^\infty \frac{\eta_i}{N^i} ~.
\end{equation}
Thus at leading order in $1/N$ the basic vertex is unique, \cite{Gracey:2}. 
Hence at this order one can integrate at either of the vertices and produce the
first contribution to the integral which is $\nu(\mu-1,\mu-1,2)$. The second 
integration is a simple chain and naively gives $\nu(\mu,\mu,0)$. This is 
clearly ill-defined due to the zeroes and singularities deriving from the 
$\Gamma$-function. However, this graph was chosen to illustrate the fact that 
the graph and indeed the theory requires a regularization in this critical
point formulation. The method developed in \cite{Gracey:1,Gracey:2} was to use 
analytic regularization which is introduced by shifting the vertex anomalous 
dimension by an infinitesimal amount, $\Delta$, via
\begin{equation}
\beta ~ \rightarrow ~ \beta ~-~ \Delta ~.
\end{equation} 
In some respect one is in effect performing a perturbative expansion in the
vertex anomalous dimension, \cite{Gracey:1,Gracey:2}. Consequently even at 
leading order the graph of Figure $10$ no longer has a unique vertex due to a 
non-zero $\Delta$. Therefore, to determine the graph to the finite part in 
$\Delta$ requires the addition and subtraction of the graphs of Figure $11$, 
\cite{Gracey:2}.
\begin{center}
\begin{picture}(230,120)(0,0)
\Line(0,50)(40,25)
\Line(80,50)(40,25)
\Line(80,50)(40,75)
\Line(40,25)(40,75)
\Vertex(0,50){1.2}
\Vertex(40,25){1.2}
\Vertex(40,75){1.2}
\Vertex(80,50){1.2}
\Text(-7,50)[]{$0$}
\Text(87,50)[]{$x$}
\Text(40,17)[]{$z$}
\Text(40,83)[]{$y$}
\Text(10,33)[]{$\alpha$}
\Text(70,33)[]{$\alpha$}
\Text(70,67)[]{$\alpha$}
\Text(57,50)[]{$\beta-\Delta$}
\Line(140,50)(180,25)
\Line(140,50)(180,75)
\Line(220,50)(180,25)
\Line(180,25)(180,75)
\Vertex(140,50){1.2}
\Vertex(180,25){1.2}
\Vertex(180,75){1.2}
\Vertex(220,50){1.2}
\Text(133,50)[]{$0$}
\Text(227,50)[]{$x$}
\Text(180,17)[]{$z$}
\Text(180,83)[]{$y$}
\Text(150,33)[]{$\alpha$}
\Text(150,67)[]{$\alpha$}
\Text(210,33)[]{$\alpha$}
\Text(197,50)[]{$\beta-\Delta$}
\end{picture}

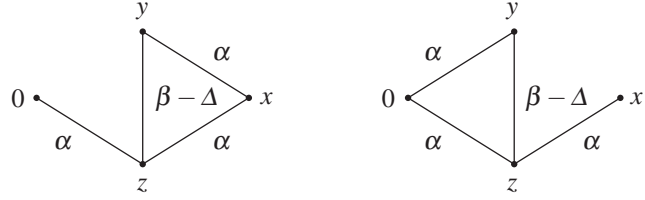
{Figure $11$. Subtracted graphs for computation of $\sigma$ two loop self 
energy.}
\end{center}
These two graphs have been chosen in such a way that their singularity
structure in $\Delta$ exactly matches that of Figure $10$, \cite{Gracey:2}. 
Clearly they represent simple chain integrals which can be determined as
$2\nu(\alpha,\beta-\Delta,2\mu-\alpha-\beta+\Delta) \nu(\alpha,\mu-\Delta,
\mu-\alpha+\Delta)$ where the singularity is clearly regularized. To complete
the evaluation introduces another technique, which we will use later, to
extract a finite term of a graph. This is a temporary regularization,
\cite{Gracey:2}. If one subtracts the graphs of Figure $11$ from that of Figure
$10$, the combination is finite with respect to $\Delta$ which is therefore not
required and can be set to zero. Thus one can complete the first integration at
the upper vertex of each graph. (Without a regularization the point where one 
integrates in each graph has to be the same and thence the order of integration
is important.) This produces $\nu(\alpha,\alpha,\beta)$ for each graph. 
However, each of the three subsequent chain integrals has a singular exponent, 
$\mu$. To circumvent this the lower two propagators of all three graphs are 
temporarily regularized by $\alpha$~$\rightarrow$~$\alpha$~$-$~$\delta$ where 
$\delta$ is arbitrary. Thus the three graphs give
\begin{eqnarray}
&& \left[ \nu(\mu-\delta,\mu-\delta,2\delta) 
- \nu(\alpha-\delta,\mu-\delta,\mu-\alpha+2\delta) \right. \nonumber \\
&& \left. ~- \nu(\mu-\delta,\alpha-\delta,\mu-\alpha+2\delta) \right]
\nu(\alpha,\alpha,\beta) 
\end{eqnarray} 
which is clearly finite as $\delta$~$\rightarrow$~$0$, \cite{Gracey:2}. Thus to
$O(\Delta)$ the graph of Figure $10$ evaluates to, \cite{Gracey:2}, 
\begin{equation}
\frac{2\pi^{2\mu} a^2(\alpha) a(\beta)}{\Gamma(\mu)} \left[ \frac{1}{\Delta}
+ B(\beta) - B(\alpha) + O(\Delta) \right]
\end{equation}
where $B(z)$~$=$~$\psi(z)$~$+$~$\psi(\mu-z)$ for $z$ and $(\mu-z)$ not equal to
zero or a negative integer and $\psi(z)$ is derivative of the logarithm of the
$\Gamma$-function.

A more involved example which uses many of the techniques of the previous
section occurs in the computation of the $O(1/N^2)$ correction to the 
$\beta$-function in $O(N)$ $\phi^4$ theory. The relevant critical exponent is
$\omega$ which is related through the critical renormalization group equation
to the $\beta$-function slope at criticality. In this case it has the form
\begin{equation}
\omega ~=~ 2 ~-~ \mu ~+~ \sum_{n=1}^\infty \frac{\omega_n}{N^n} 
\end{equation}
and the explicit forms for $\omega_n$ are deduced from the part of the
Schwinger-Dyson equations corresponding to corrections to scaling. In other
words the propagators of (\ref{Gracey:propdef}) are extended to
\begin{eqnarray}
\langle \phi^i(0) \phi^j(x) \rangle &=& \frac{\delta^{ij}A}{(x^2)^\alpha} 
\left[ 1 ~+~ A^\prime (x^2)^\omega \right] \nonumber \\
\langle \sigma(0) \sigma(x) \rangle &=& \frac{B}{(x^2)^\beta} 
\left[ 1 ~+~ B^\prime (x^2)^\omega \right] ~. 
\end{eqnarray}
In principle other corrections can appear here corresponding to other exponents
such as that for the $\beta$-function of the nonlinear $\sigma$ model but one 
tends to focus on one calculation at a time. The effect of the corrections is 
that to deduce $\omega_n$ within the Schwinger-Dyson formalism all Feynman 
diagrams with {\em one} correction insertion on a propagator contribute at each
particular order in $1/N$. While the $O(1/N^2)$ expression for $\omega$ 
appeared in \cite{Gracey:15} the explicit evaluation of the contributing graphs
has not been detailed. Thus we discuss one such diagram here as the approach 
can be readily adapted to the other graphs. It is given in Figure $12$. To see
that it is $O(1/N^2)$ each closed loop of $\phi^i$ fields contributes a factor
of $N$ and each $\sigma$ propagator is $O(1/N)$. This is due to the fact that
the amplitude $B$ is $O(1/N)$, \cite{Gracey:1,Gracey:2}. As there are four of
the former and five of the latter then this gives $O(1/N)$ overall which is one
factor of $1/N$ more than the previous order graph of Figure $10$. Finally,
another factor of $1/N$ derives from the actual Schwinger Dyson formalism used
to determine $\omega_2$. The double line on one $\sigma$ propagator in Figure
$12$ denotes the $B^\prime$ correction. 
\begin{center}
\begin{picture}(320,200)(0,-15)
\Gluon(0,80)(30,80)53
\Line(30,80)(80,20)
\Line(30,80)(80,140)
\Line(80,20)(80,140)
\Gluon(80,20)(130,20)55
\Gluon(80,140)(130,140)55
\Line(104,130)(104,150)
\Line(106,130)(106,150)
\Line(130,140)(180,140)
\Line(130,140)(155,100)
\Line(155,100)(180,140)
\Line(130,20)(180,20)
\Line(130,20)(155,60)
\Line(155,60)(180,20)
\Gluon(155,60)(155,100)53
\Gluon(180,20)(230,20)55
\Gluon(180,140)(230,140)55
\Line(230,20)(230,140)
\Line(230,20)(280,80)
\Line(230,140)(280,80)
\Gluon(280,80)(310,80)53
\end{picture}

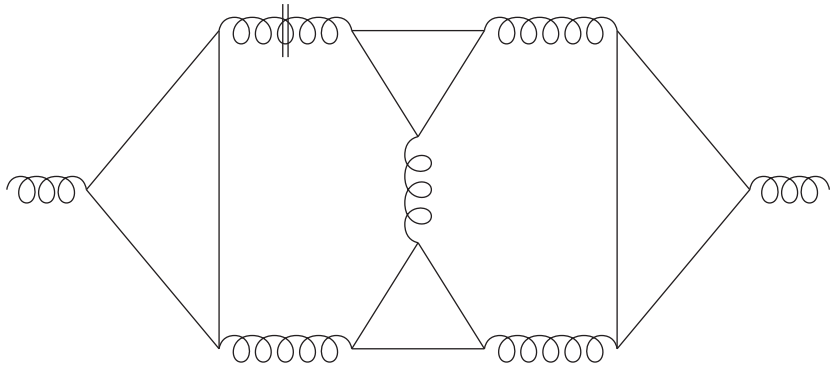
{Figure $12$. Particular graph contributing to the $\phi^4$ theory 
$\beta$-function at $O(1/N^2)$.}
\end{center}
The presence of such a correction means that the graph is $\Delta$-finite.
Moreover, since we only want the value as a function of $d$ rather than $d$ 
and $N$ we can replace the exponents of the lines by their canonical values. If
one was computing $\omega_3$ then the anomalous dimensions of each exponent 
would need to be retained at $O(1/N)$. The benefit of this restriction here is 
that of the ten vertices eight are unique. There are ten integrations to do 
over the vertices rather than the six of the loops as we are in coordinate 
space representation. Given this high degree of uniqueness the graph can be 
reduced rather quickly to one with fewer integrations. To do this one can use a
variety of the rules we had earlier aside from uniqueness such as conformal
transformation, unique triangle, insertion at an internal or external vertex.
Ultimately one produces the graph of Figure $13$.
\begin{center}
\begin{picture}(210,180)(0,-25)
\Line(10,80)(50,130)
\Line(10,80)(100,0)
\Line(50,130)(100,0)
\Line(100,0)(150,130)
\Line(50,130)(150,130)
\Line(150,130)(190,80)
\Line(100,0)(190,80)
\Vertex(10,80){1.2}
\Vertex(50,130){1.2}
\Vertex(100,0){1.2}
\Vertex(150,130){1.2}
\Vertex(190,80){1.2}
\Text(100,140)[]{$1$}
\Text(160,30)[]{$1$}
\Text(30,30)[]{$\mu-1$}
\Text(145,80)[]{$1$}
\Text(50,80)[]{$\mu-1$}
\Text(10,110)[]{$\mu-1$}
\Text(190,110)[]{$2\mu-3$}
\end{picture}

{Figure $13$. Reduced integral of Figure $12$.}
\end{center}
This graph cannot be reduced any further since there are no unique vertices or
triangles. Though various vertices or triangles are one unit from uniqueness.
Moreover, integration by parts cannot be used since at some point one produces
an unregularized exponent, such as $0$ or $\mu$, or a zero in a denominator 
factor. In some sense this graph could be regarded as a master integral since 
it arises in several of the other graphs contributing to the $\sigma$ 
Schwinger-Dyson equation. Moreover, it is worth noting that in strictly four 
dimensions the propagators of the graph would all have unit exponents. As an 
aside if an interested reader has been applying the conformal techniques to 
reduce the diagram and obtains similar exponents but distributed differently 
around the diagram then it will be related to that of Figure $13$ by applying 
the transformations discussed for the master $2$-loop self energy. We note here
that if a conformal transformation is applied to the graph of Figure $13$ with 
the left internal point as the CL base, then that would introduce a new line 
from the top right internal vertex to the base. This illustrates comments made 
earlier.

To proceed further and reduce the graph to a known function of $d$ requires an
integration by parts but this requires modifying the integral first. Though
before this can be achieved safely one has to introduce a temporary 
regularization to handle hidden singularities at a later stage of the 
computation. This technique has been applied by others, 
\cite{Gracey:4,Gracey:16}. For our case we have choosen the regularization of 
Figure $14$. How one chooses the temporary regularization is not unique. 
However, it is chosen here so that after application of the integration by 
parts rule of Figure $7$ the resulting four graphs have either unique vertices 
or triangles which are $\delta$-dependent and which regularize any singularity 
after subsequent integration. For the integration by parts we use the top left 
\begin{center}
\begin{picture}(210,170)(0,-10)
\Line(10,80)(50,130)
\Line(10,80)(100,0)
\Line(50,130)(100,0)
\Line(100,0)(150,130)
\Line(50,130)(150,130)
\Line(150,130)(190,80)
\Line(100,0)(190,80)
\Vertex(10,80){1.2}
\Vertex(50,130){1.2}
\Vertex(100,0){1.2}
\Vertex(150,130){1.2}
\Vertex(190,80){1.2}
\Text(100,140)[]{$1$}
\Text(170,30)[]{$1+\delta$}
\Text(20,30)[]{$\mu-1-\delta$}
\Text(145,80)[]{$1-\delta$}
\Text(40,80)[]{$\mu-1+\delta$}
\Text(0,110)[]{$\mu-1-\delta$}
\Text(200,110)[]{$2\mu-3+\delta$}
\end{picture}

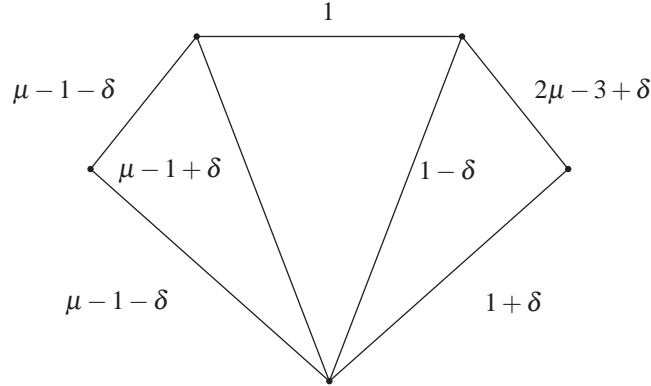
{Figure $14$. Temporary regularization of previous graph to reduce it to two
loop basic graphs.}
\end{center}
internal vertex of Figure $14$ with the line joining the quartic vertex as the 
reference line of the rule of Figure $7$. This produces the four graphs of 
Figures $15$-$18$.
\begin{center}
\begin{picture}(210,160)(0,-10)
\Line(10,80)(50,130)
\Line(10,80)(100,0)
\Line(50,130)(100,0)
\Line(100,0)(150,130)
\Line(50,130)(150,130)
\Line(150,130)(190,80)
\Line(100,0)(190,80)
\Vertex(10,80){1.2}
\Vertex(50,130){1.2}
\Vertex(100,0){1.2}
\Vertex(150,130){1.2}
\Vertex(190,80){1.2}
\Text(100,140)[]{$2$}
\Text(170,30)[]{$1+\delta$}
\Text(20,30)[]{$\mu-1-\delta$}
\Text(145,80)[]{$1-\delta$}
\Text(40,80)[]{$\mu-2+\delta$}
\Text(0,110)[]{$\mu-1-\delta$}
\Text(200,110)[]{$2\mu-3+\delta$}
\end{picture}

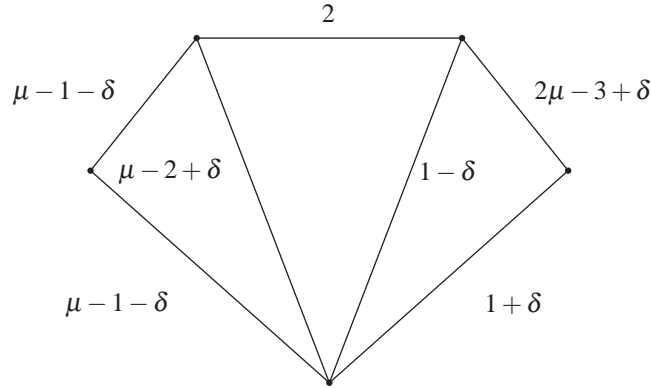
{Figure $15$. First graph after integration by parts.}
\end{center}
All but the third have at least one unique vertex while that has a unique 
triangle. In our earlier notation the first two graphs of Figure $15$ and $16$
are
\begin{equation}
\nu(2,1-\delta,2\mu-3+\delta)
I(\mu-1-\delta,\mu-1+\delta,\mu-1+\delta,\mu-1-\delta,1)
\end{equation}
and
\begin{equation}
\nu(2,\mu-1-\delta,\mu-1+\delta)
I(1+\delta,2\mu-3,1+\delta,2\mu-3-\delta,1) ~.
\end{equation}
As both of these are $\delta$-finite and have no $\delta$-singular
coefficients, one can set $\delta$ to zero in each. The final evaluation is by
a two loop reduction formula similar to those of Figures $8$ and $9$.
\begin{center}
\begin{picture}(210,170)(0,-20)
\Line(10,80)(50,130)
\Line(10,80)(100,0)
\Line(50,130)(100,0)
\Line(100,0)(150,130)
\Line(50,130)(150,130)
\Line(150,130)(190,80)
\Line(100,0)(190,80)
\Vertex(10,80){1.2}
\Vertex(50,130){1.2}
\Vertex(100,0){1.2}
\Vertex(150,130){1.2}
\Vertex(190,80){1.2}
\Text(100,140)[]{$2$}
\Text(170,30)[]{$1+\delta$}
\Text(20,30)[]{$\mu-1-\delta$}
\Text(145,80)[]{$-\delta$}
\Text(40,80)[]{$\mu-1+\delta$}
\Text(0,110)[]{$\mu-1-\delta$}
\Text(200,110)[]{$2\mu-3+\delta$}
\end{picture}

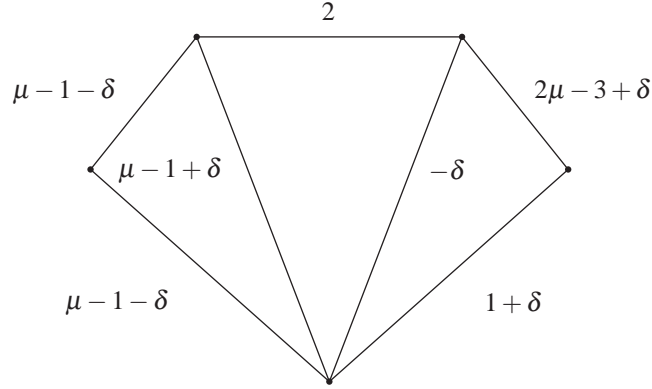
{Figure $16$. Second graph after integration by parts.}
\end{center}
For the remaining two graphs of Figures $17$ and $18$ one has to treat them
together due to the singular propagator exponents as will be evident. After 
integrating the respective unique triangle and vertex they combine to produce
\begin{eqnarray}
&& a^3(1) a(\mu-\delta) a(2\mu-3+\delta) \left[ 
I(\mu-1,\mu-1,1+\delta,\mu-1-\delta,\mu-1 \right. \nonumber \\
&& \left. ~~~~~~~~~~~~~~~~~~~~~~~~~~~~~~-~
I(\mu-1,\mu-1,1+\delta,\mu-1-2\delta,\mu-1+\delta) \right] ~.
\end{eqnarray}
\begin{center}
\begin{picture}(210,170)(0,-20)
\Line(10,80)(50,130)
\Line(10,80)(100,0)
\Line(50,130)(100,0)
\Line(100,0)(150,130)
\Line(50,130)(150,130)
\Line(150,130)(190,80)
\Line(100,0)(190,80)
\Vertex(10,80){1.2}
\Vertex(50,130){1.2}
\Vertex(100,0){1.2}
\Vertex(150,130){1.2}
\Vertex(190,80){1.2}
\Text(100,140)[]{$1$}
\Text(170,30)[]{$1+\delta$}
\Text(20,30)[]{$\mu-1-\delta$}
\Text(145,80)[]{$1-\delta$}
\Text(40,80)[]{$\mu-2+\delta$}
\Text(0,110)[]{$\mu-\delta$}
\Text(200,110)[]{$2\mu-3+\delta$}
\end{picture}

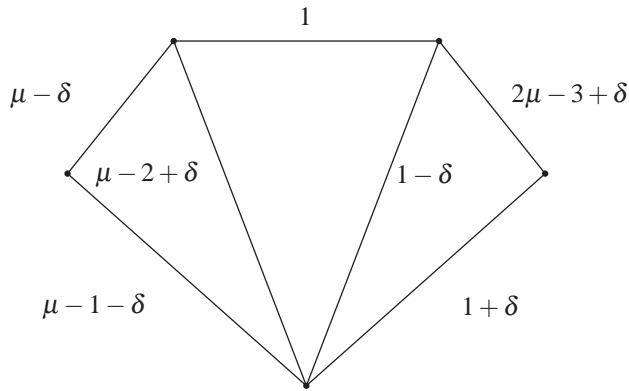
{Figure $17$. Third graph after integration by parts.}
\end{center}
As the external coefficient includes a factor of $1/\delta$ then the quantity
inside the square brackets needs to be evaluated to $O(\delta)$. 
\begin{center}
\begin{picture}(210,170)(0,-10)
\Line(10,80)(50,130)
\Line(10,80)(100,0)
\Line(50,130)(100,0)
\Line(100,0)(150,130)
\Line(50,130)(150,130)
\Line(150,130)(190,80)
\Line(100,0)(190,80)
\Vertex(10,80){1.2}
\Vertex(50,130){1.2}
\Vertex(100,0){1.2}
\Vertex(150,130){1.2}
\Vertex(190,80){1.2}
\Text(100,140)[]{$1$}
\Text(170,30)[]{$1+\delta$}
\Text(20,30)[]{$\mu-2-\delta$}
\Text(145,80)[]{$1-\delta$}
\Text(40,80)[]{$\mu-1+\delta$}
\Text(0,110)[]{$\mu-\delta$}
\Text(200,110)[]{$2\mu-3+\delta$}
\end{picture}

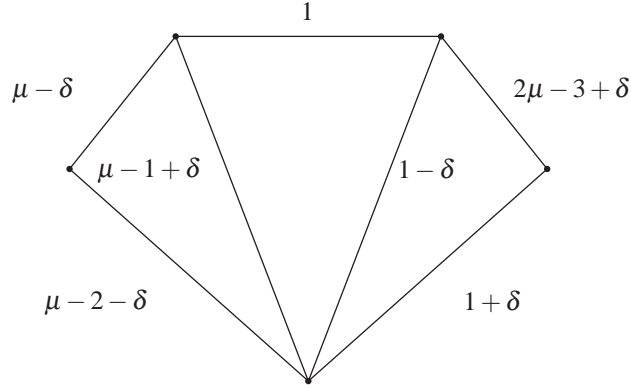
{Figure $18$. Fourth graph after integration by parts.}
\end{center}
This is not possible exactly for both integrals. (It is for the first.) Instead
since one only needs the $O(\delta)$ part itself one can achieve this by 
evaluating the integral
\begin{equation}
I(\mu-1,\mu-1,1,\mu-1-\delta,\mu-1+\delta) ~.
\end{equation}
From the two $2$-loop graphs we are interested in the $O(\delta)$ term of this 
integral clearly corresponds to the piece we require. Moreover, it can be
evaluated exactly using $\downarrow$ as it then reduces to an integral to which
one can apply a $2$-loop recurrence relation similar to that of Figure $9$. The
final expression for the graph of Figure $12$ is
\begin{equation}
\frac{a(\mu-1) a^2(2\mu-3) a(2\mu-2)}{2(\mu-3)(\mu-2)^9} \left[ f_2 - f_1^2 
- \frac{2f_1}{(\mu-2)} + 6 f_3 \right]
\end{equation}
where
\begin{eqnarray}
f_1 &=& \psi(3-\mu) + \psi(2\mu-3) - \psi(\mu-1) - \psi(1) \nonumber \\
f_2 &=& \psi^\prime(3-\mu) - \psi^\prime(2\mu-3) + \psi^\prime(\mu-1) 
- \psi^\prime(1) \nonumber \\
f_3 &=& \psi^\prime(\mu-1) - \psi^\prime(1) ~. 
\end{eqnarray}
Setting $\mu$~$=$~$2$ reproduces the established leading order value for the 
wheel of three spokes, \cite{Gracey:17}, which provides a useful check. 
Finally, all the other contributing graphs are evaluated in a similar way and
the full expression for $\omega_2$, after using the Schwinger Dyson formalism, 
is given in \cite{Gracey:15}.

\section{Future Directions}

We close the article by discussing several directions in which this approach
could move. First, the extension of scalar field theories to non-abelian gauge
theories has been considered in 
\cite{Gracey:18,Gracey:19,Gracey:20,Gracey:21,Gracey:22} for various  
applications where information is needed on the renormalization group functions
of operators in deep inelastic scattering and the $\beta$-function. That
approach is based on the observations of \cite{Gracey:23} using the number of 
quark flavours, $\Nf$, as the expansion parameter. Rather than use the full QCD
Lagrangian one exploits the critical point equivalence with the non-abelian
Thirring model, \cite{Gracey:23}, 
\begin{equation}
L ~=~ i \bar{\psi}^i \Dslash \psi^i - \half (A^a_\mu)^2
\label{Gracey:natmlag}
\end{equation}
where $D_\mu$ is the covariant derivative, $T^a$ are the group generators and 
$\psi^i$ is the quark field with $1$~$\leq$~$i$~$\leq$~$\Nf$. The spin-$1$ 
auxiliary field $A^a_\mu$ plays the role of the gluon in the higher dimensional
theory. The triple and quartic gluon vertices of QCD are generated by the 
$3$-point and $4$-point functions of (\ref{Gracey:natmlag}) with $A^a_\mu$ 
external legs respectively. Following the critical point analysis the 
propagators in a similar notation, but in momentum space, are
\begin{eqnarray}
\langle \psi^i(-p) \psi^j(p) \rangle &=& 
\frac{\delta^{ij}A\pslash}{(p^2)^{\mu-\alpha}} \nonumber \\
\langle A^a_\mu(-p) A^b_\nu(p) \rangle &=& -~ 
\frac{\delta^{ab}B}{(p^2)^{\mu-\beta}} \left[ \eta_{\mu\nu} - \xi 
\frac{p_\mu p_\nu}{p^2} \right]
\end{eqnarray}
where $\xi$ is the gauge parameter with the Landau gauge corresponding to
$\xi$~$=$~$1$. From dimensional analysis the exponents are now, 
\cite{Gracey:18},
\begin{equation}
\alpha ~=~ \mu ~+~ \half \eta ~~~~,~~~~ \beta ~=~ 1 ~-~ \eta ~-~ \chi
\end{equation}
which means the basic vertex is one step from uniqueness. This complicates
computations in that to proceed one has to break all contributing graphs into
scalar integrals and treat them by transformations, subtractions or use
integration by parts to reduce them to computable cases. While it has been 
possible to do this in certain instances, \cite{Gracey:19,Gracey:20,Gracey:22},
it is not systematic. 

Since the application of the method of \cite{Gracey:1,Gracey:2} to QCD an 
algorithm has been developed which allows one to exploit integration by parts. 
Known as the Laporta\index{Laporta} algorithm, \cite{Gracey:11}, it creates all
integration by parts relations between integrals of a particular topology and 
then algebraically solves them in such a way that all integrals are reduced to 
a basis set of master integrals. Once their values are known by other methods 
then the problem is complete. In the large $\Nf$ context once one moves to say 
$O(1/\Nf^2)$ computations then graphs such as that of Figure $12$ need to be 
computed in QCD. Then the solid lines would represent quarks and the springs 
would correspond to gluons. However, taking the traces over the closed loops 
results in a huge number of irreducible numerator scalar products. While the 
propagators do not have integer powers, as is the case in perturbative 
calculations, there appears to be a similarity to the problem. In other words 
in principle a generalization of the Laporta\index{Laporta} algorithm should be
able to produce a reduction of the irreducible graphs to a set of masters. The 
difficulty is that the presence of non-integer propagator powers means that the
present Laporta\index{Laporta} algorithm would need to be modified in order to 
have a point, akin to a ground state, below which no more reductions could be 
possible. Though it is not clear under what conditions such a bottom point 
exists or whether for certain topologies or distribution of non-unit exponents 
it can be proved to be impossible. Indeed the latter point could be related to 
the issue of lack computability of a graph due to the presence of multiple-zeta
values \index{multiple zeta} similar to (\ref{Gracey:intdjb}). However, it 
seems that for the practical problem of deducing the QCD $\beta$-function at 
$O(1/\Nf^2)$ such an extension to the Laporta algorithm is possibly the only 
feasible tactic at present.

Aside from this possible extension to the Laporta\index{Laporta} algorithm
another interesting possibility is to what extent the 
conformal\index{conformal} integration methods can be built into that algorithm
to improve and speed reductions within a computer algebra programme for 
massless Feynman graphs. This may be important for higher loop topologies. For 
instance, earlier we derived recurrence relations for the two loop self energy 
topology based on the transformation deduced from the generalized uniqueness 
condition. While such relations are no doubt contained within integration by 
parts relations of the Laporta construction, that of Figure $9$ is particularly
useful in that there is no increase in the power of any propagator. Therefore, 
it may be possible to construct similar relations using 
conformal\index{conformal} transformations but for higher loop massless 
topologies. Indeed such transformations are not unrelated to the symmetry group
of the topology as has already been studied in depth for the two loop self 
energy, \cite{Gracey:6,Gracey:7}. At the time of \cite{Gracey:6,Gracey:7} 
expanding a graph in terms of its group invariants was a promising approach 
which was complemented by later methods such as \cite{Gracey:8,Gracey:9}. 
However, it may be worth returning to a group theory analysis for topologies 
such as that represented in Figure $13$. This is because the high order 
expansion in terms of $\epsilon$ of this and other three and four loop 
topologies will soon be required for extending QCD to {\em five} and possibly 
higher loops. In this respect another direction of exploration may be to study 
the structure of the graph polynomials of a topology. The transformations of 
\cite{Gracey:2} have been derived from a graphical approach to understanding 
the structure of the two loop self energy graph. Understanding the effect such 
conformal transformations have on the graph polynomials of massless integrals 
may also give insight into the as yet undetermined group theory properties of 
higher order topologies.  

\begin{acknowledgement} It is with pleasure that I thank the organizers of the
meeting for permission to include this article in these proceedings. It is 
based on a talk presented at {\em Quantum Field Theory, Periods and 
Polylogarithms III}, Humboldt University, Berlin in June 2012 which was also in
honour of Dr D.J. Broadhurst's $65$th birthday. The {\sc Axodraw} package,
\cite{Gracey:24}, was used to draw the figures in the article.
\end{acknowledgement}

\end{document}